%% file: CSCloud2027_5.tex
\documentclass[conference]{IEEEtran}
\IEEEoverridecommandlockouts

\usepackage{amsmath,amssymb,amsthm}
\usepackage{booktabs}
\usepackage{graphicx}
\usepackage{microtype}
\usepackage{cite}
\usepackage{url}
\usepackage{xcolor}

\graphicspath{{./}}
\IfFileExists{results_macros.tex}
  {\input{results_macros.tex}}
  {\newcommand{\OursMissPct}{TBD}
   \newcommand{\OursMissLoPct}{TBD}
   \newcommand{\OursMissHiPct}{TBD}
   \newcommand{\OursEnergy}{TBD}
   \newcommand{\FixedSavingPct}{TBD}
   \newcommand{\HardSavingPct}{TBD}
   \newcommand{\PooledMissPct}{TBD}
   \newcommand{\MedianMissPct}{TBD}
   \newcommand{\MeanMissPct}{TBD}
   \newcommand{\RuntimeP}{TBD}
   \newcommand{\ExactGapPct}{TBD}
   \newcommand{\ExactRuntimeRatio}{TBD}
   \newcommand{\CleanOverheadPct}{TBD}
   \newcommand{\UnusedCapEnergy}{TBD}
   \newcommand{\OracleEnergyGapPct}{TBD}
   \newcommand{\OracleMissPct}{TBD}
   \newcommand{\WorstcaseEnergy}{TBD}
   \newcommand{\WorstcaseMissPct}{TBD}
   \newcommand{\DenialMissPct}{TBD}
   \newcommand{\DenialDropPct}{TBD}
   \newcommand{\NumSeeds}{TBD}
   \newcommand{\NumBatches}{TBD}}

\newcommand{\method}{FleetGuard}

\newtheorem{theorem}{Theorem}

\title{\method: Identity-Capped Conformal Risk Allocation for Reliable UAV Edge--Cloud Offloading}

\author{
\IEEEauthorblockN{\parbox{\textwidth}{\centering
Junhao Wei$^{1}$,\hspace{0.1em}Haochen Li$^{1}$,\hspace{0.1em}
Yifu Zhao$^{1}$,\hspace{0.1em}Yanxiao Li$^{1}$,\hspace{0.1em}
Dexing Yao$^{1}$,\hspace{0.1em}Baili Lu$^{1}$,\\
\mbox{Ngai Cheong$^{1}$,~\IEEEmembership{Senior Member, IEEE}},
Sio-Kei Im$^{2}$,~\IEEEmembership{Senior Member, IEEE},\\
Xu Yang$^{1}$,~\IEEEmembership{Member, IEEE}, and
Yapeng Wang$^{1,*}$,~\IEEEmembership{Member, IEEE}}}
\IEEEauthorblockA{$^{1}$Faculty of Applied Sciences,
Macao Polytechnic University, Macao 999078, China}
\IEEEauthorblockA{$^{2}$Macao Polytechnic University,
Macao 999078, China}
\thanks{$^{*}$Corresponding author: Yapeng Wang
(yapengwang@mpu.edu.mo). This work (MPU submission code: fca.66d9.0e53.1
) is supported by the grant from Macao
Polytechnic University (RP/FCA-06/2026) and the Macao Science and Technology
Development Fund (FDCT-MOST: 0018/2025/AMJ).}
}

\begin{document}
\maketitle

\begin{abstract}
Compromised UAV reporters can poison latency telemetry and attract offloaded
tasks to degraded servers.  We present \method, an identity-capped conformal
offloader for local, edge, and cloud execution.  With equal buffers of \(m\)
scores and a cap of \(K\) compromised identities, a \(Km\) rank correction
retains one-sided fixed-level coverage without an honest-reporter majority.
A deterministic repair scheduler allocates task-specific risk under fleet,
capacity, and deadline constraints.  Heavy-tailed simulations with eight
reporters, four edges, \NumSeeds{} independent seeds, and \NumBatches{}
batches per seed and condition yield \OursMissPct\% deadline misses under
seven low-score compromised reporters (95\% clustered interval:
\OursMissLoPct--\OursMissHiPct\%), compared with approximately 9.6\% for
pooled conformal, reporter-median, and mean rules.  \method{} saves
\FixedSavingPct\% energy over fixed-risk BRC and approaches adaptive exact
assignment within \ExactGapPct\%, with \ExactRuntimeRatio\(\times\) lower
mean scheduling time.
\end{abstract}


\section{Introduction}

Unmanned aerial vehicles (UAVs) use edge servers and clouds for
latency-sensitive perception and planning \cite{Li2026RAWARHP}, including self-attention-based
navigation \cite{wei2026saga} and vision-language mission generation
\cite{Wei2026ARIESMission} \cite{wei2026aries2}.  Offloading reduces onboard energy when remote execution
meets task deadlines.  Multi-UAV controllers jointly allocate placement,
communication, computation, and trajectory
\cite{yang2020loadbalance}\cite{guo2024multiuav}\cite{xu2024trajectory}; their decisions
depend on reliable service-time tail estimates.

Pooling telemetry across authenticated UAVs improves tail estimation, but
compromised identities can submit small latency residuals and make a degraded
edge appear safe.  This attack poisons scheduling evidence while communication
remains available.  It differs from jamming \cite{liu2024jamming},
eavesdropping \cite{cui2025robust}, and the sparse operational faults detected
by flight-log anomaly mining \cite{wei2026aerotsboost}.

Conformal prediction provides finite-sample marginal coverage under
exchangeability \cite{angelopoulos2023gentle}, but ordinary pooling loses
this property under arbitrary score poisoning.  The closest calibration
studies, Rob-FCP \cite{kang2024robfcp} and its partial-sharing extension
\cite{lari2026partial}, filter anomalous client sketches.  Fleet controllers
can enforce a different trust mechanism: equal per-identity score caps.
FleetGuard uses these caps to certify a padded order rank even without an
honest-reporter majority.

The certified bound curve then supports task-specific risk allocation.
Deadline, edge-capacity, and fleet-risk constraints couple the execution
choices; fixed-risk protection can waste energy.  We combine the certificate
with deterministic repair and compare its energy cost with a matched optimal
assignment.

The main contributions are:
\begin{itemize}
  \item An identity-capped conformal rank certificate with fixed-level
  coverage without an honest-reporter majority.
  \item A deterministic scheduler for joint destination and risk allocation
  under deadline, capacity, and fleet-risk constraints.
  \item Paired simulations demonstrating \FixedSavingPct\% energy savings
  over fixed BRC and a \ExactGapPct\% optimality gap.
\end{itemize}

\section{Related Work}

\subsection{UAV Computing and Resource Allocation}

UAV edge computing combines load-balanced assignment \cite{yang2020loadbalance},
cooperative offloading \cite{guo2024multiuav}, reliability-aware altitude and
offloading design \cite{hui2024altitude}, and multi-timescale trajectory and
resource allocation \cite{xu2024trajectory}.  Related systems allocate
resources for federated learning \cite{liu2023federated}, emergency offloading
\cite{chen2023emergency}, and bandwidth-constrained edge dispatch
\cite{zhang2019dedas}.  Adaptive particle swarm optimization
\cite{wei2024path} and geometric sparrow search \cite{wei2026geossa}
optimize UAV paths, while LA-BHH selects search operators for inspection
routing \cite{wei2026labhh}.  These studies address computation or motion;
FleetGuard protects the latency calibration used for task execution.

Security-aware offloading optimizes communication against jamming
\cite{liu2024jamming} and uncertain tasks with aerial eavesdroppers
\cite{cui2025robust}.  Our adversary instead replaces held-out residual
scores after latency-model training.

\subsection{Robust Conformal Calibration}

Split conformal calibration uses held-out order statistics
\cite{angelopoulos2023gentle}; shift, dependence, and label noise affect its
coverage \cite{barber2023beyond}cite{einbinder2024noise}.  Poisoning can manipulate
conformal uncertainty \cite{li2024poison}.  RPS aggregates prediction sets from
calibration partitions \cite{scholten2025rps}, and ACon\(^{2}\) constructs
online consensus sets under shift and Byzantine oracles \cite{park2023acon2}.
Rob-FCP characterizes client distributions through sketches and filters
anomalous clients using benign-client references \cite{kang2024robfcp}.
Its partial-sharing extension protects training and calibration while
retaining histogram-based characterization and client exclusion
\cite{lari2026partial}.

FleetGuard authenticates contributors, enforces equal scalar-score buffers,
and uses a declared compromise cap \(K\).  These ingestion controls permit a
direct fixed-level rank certificate at \(K=U-1\), without identifying malicious
reporters.  The resulting latency-bound curve enters a capacity-constrained
fleet assignment problem.

\section{System and Threat Model}

\begin{figure*}[t]
  \centering
  \includegraphics[width=\textwidth]{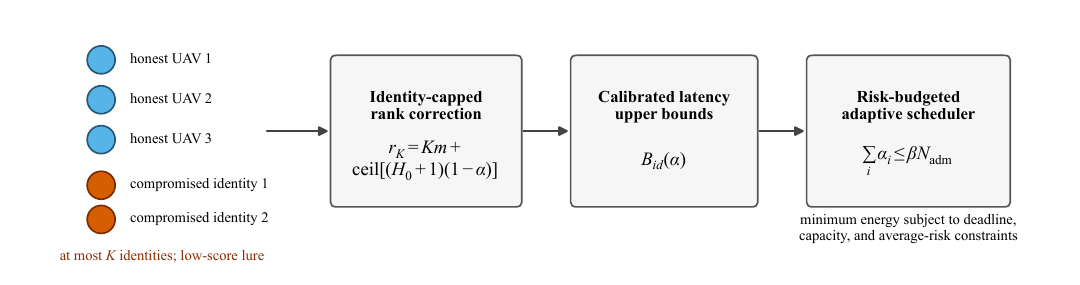}
  \caption{Trust boundary and scheduling path.  Authenticated reporters
  contribute equal buffers.  Rank padding yields latency bounds for
  energy-minimizing scheduling under deadline, capacity, and nominal-risk
  constraints.  \(N_{\rm adm}\) counts admitted tasks; ceil rounds upward.}
  \label{fig:system}
\end{figure*}

Fig.~\ref{fig:system} presents the trust boundary and scheduling path.  A
decision epoch contains \(N\) tasks and destinations
\(\mathcal D=\{0,1,\ldots,D,D+1\}\).  Index 0 denotes local execution,
indices \(1,\ldots,D\) denote ground-edge servers, and index \(D+1\) denotes a
remote cloud.  The scheduler operates after task generation and before remote
service begins.

\subsection{Task and Service Model}

Task \(i\in\{1,\ldots,N\}\) has deadline \(T_i\) in seconds,
destination-dependent execution energy \(e_{id}\) in joules, and edge demand
\(c_{id}\) in Gcycles.  Let \(L_{id}\) denote the realized latency in seconds.
A latency predictor, fixed before calibration, provides a predicted latency
\(\widehat L_{id}\) and a positive residual scale
\(\widehat\sigma_{id}\), both in seconds, for each remote destination.  The normalized
one-sided score is
\begin{equation}
  S_{id}=\frac{L_{id}-\widehat L_{id}}{\widehat\sigma_{id}}.
  \label{eq:score}
\end{equation}
The dimensionless score allows tasks with different predicted latencies and
scales to share destination-specific calibration buffers under the
exchangeability assumption below.  Local execution is
deterministic in the present model.  Remote execution uses calibrated upper
bounds.  A deadline miss includes either late completion or admission failure.

The energy objective includes UAV computation and transmission plus allocated
gateway and backhaul energy for cloud execution.  Edge capacities model batch
compute contention; aircraft positions remain fixed within a batch.

\subsection{Telemetry Ingestion Assumptions}

For each remote destination, the controller accepts exactly \(m\) scores from
each of \(U\) authenticated identities during a calibration epoch.  Admission
control and rate limiting prevent one identity from contributing more than
\(m\) observations.  At most \(K<U\) identities are compromised.  Scores from
honest identities and the future test score are exchangeable after
normalization.  Calibration and evaluation streams are disjoint.

The eligible identities and their contribution counts are fixed before
calibration.  Longer buffers are subsampled to \(m\) scores using a
score-independent rule; shorter buffers require a smaller common \(m\) or a
new eligible set with updated \(U\) and \(K\).  Selecting or excluding
identities based on observed honest scores would require a separate
exchangeability argument.

\subsection{Adversary and Security Objective}

A compromised identity can replace every score in its buffer by an arbitrary
finite value.  Small reports form the coverage-critical lure direction.  They
shrink an ordinary upper quantile and attract tasks to a degraded edge.  Large
reports inflate upper bounds and can remove remote options.  Mixed attacks
combine both directions across destinations or identities.

The security objective is one-sided statistical coverage for every fixed
risk level under at most \(K\) compromised buffers.  BRC bounds remain
conservative under high reports, which can suppress remote execution options.  Task drops are counted as deadline misses so that
availability loss remains visible in end-to-end evaluation.

\section{Proposed Method}

\subsection{Identity-Capped Rank Certificate}

For one remote destination, let \(\alpha\in(0,1)\) be the nominal
miscoverage level, \(p=1-\alpha\), and \(H_0=(U-K)m\) the minimum number of
honest scores.  Destination subscripts are omitted in the rank derivation.  The padded pooled rank and corresponding score are
\begin{equation}
 r_K=Km+\left\lceil(H_0+1)p\right\rceil,
 \qquad
 \widehat q_K(\alpha)=S^{\mathrm{pool}}_{(r_K)}.
 \label{eq:rank}
\end{equation}
Here \(S^{\mathrm{pool}}_{(r)}\) is the \(r\)-th smallest of the \(Um\)
reported scores, using one-based ranks.  We set \(\widehat q_K(\alpha)=+\infty\)
when \(r_K>Um\).  Otherwise, the BRC latency upper bound
for task \(i\) and destination \(d\) is
\begin{equation}
 B_{id}(\alpha)=
 \widehat L_{id}+\widehat\sigma_{id}\widehat q_{K,d}(\alpha).
 \label{eq:latencybound}
\end{equation}
The additive \(Km\) term reserves every potentially compromised score below
the selected order statistic.  It does not estimate the identities or inspect
their reported values.

\begin{theorem}[Fixed-level coverage]
Under the authentication, equal-buffer, and exchangeability assumptions,
any \(\alpha\in(0,1)\) fixed independently of the calibration scores and any
reports from at most \(K\) compromised
identities satisfy
\begin{equation}
 \Pr\!\left\{S_{\mathrm{test}}\leq\widehat q_K(\alpha)\right\}
 \geq1-\alpha.
 \label{eq:coverage}
\end{equation}
\end{theorem}

\begin{proof}
If \(r_K>Um\), the infinite bound covers every finite test score.
Otherwise, suppose the actual compromise count is \(k\leq K\).  The pooled buffer contains
\(H=(U-k)m=H_0+\delta\) honest scores, where
\(\delta=(K-k)m\).  At most \(km\) corrupt scores can occur below the selected
pooled order statistic.  Consequently,
\begin{equation}
\widehat q_K(\alpha)\geq
S^{\mathrm{hon}}_{\left(\delta+
\lceil(H_0+1)p\rceil\right)}.
\label{eq:honest_rank}
\end{equation}
Since \(0\leq p\leq1\),
\begin{equation}
\delta+\lceil(H_0+1)p\rceil
\geq\lceil(H_0+\delta+1)p\rceil
=\lceil(H+1)p\rceil.
\end{equation}
Here \(S^{\mathrm{hon}}_{(j)}\) is the \(j\)-th smallest honest score.
The rightmost rank is the finite-sample split-conformal rank of the honest
scores, with an infinite bound if that rank exceeds \(H\).  Exchangeability yields~\eqref{eq:coverage}.  Corrupt scores above the
selected pooled rank can only increase the bound.
\end{proof}

The certificate covers the realized task latency at each fixed level:
\(\Pr\{L_{id}\leq B_{id}(\alpha)\}\geq1-\alpha\), because the residual
scale is positive.  It remains valid at \(K=U-1\), with one honest buffer.
A finite bound exists exactly when \(\alpha\geq1/(H_0+1)\).
For example, \(U=8\), \(K=7\), \(m=10{,}000\), and \(\alpha=0.05\)
give \(r_K=79{,}501\): seven low poisoned buffers occupy the first
70,000 positions, leaving the 9,501st honest score as the bound.

\subsection{Certified Option Construction}

The scheduler evaluates a finite grid \(\mathcal A\) of nominal risk levels.
For a remote destination, an option \(o=(i,d,\alpha)\), with
\(\alpha\in\mathcal A\), is retained only when
\begin{equation}
 B_{id}(\alpha)\leq T_i.
 \label{eq:feasible}
\end{equation}
Energy and edge demand do not depend on \(\alpha\) after \(i\) and \(d\) are
fixed.  Therefore, only the smallest feasible risk level for each
task--destination pair is Pareto-efficient.  Larger feasible levels consume
more fleet risk without reducing energy or resource use.  This dominance rule
reduces the option set to at most one calibrated option per remote
task--destination pair.

A local option is retained only if \(L_{i0}=\widehat L_{i0}\leq T_i\);
it receives zero nominal risk because local latency is deterministic.  Every task also receives an explicit drop
option.  The drop option carries zero execution energy, a large lexicographic
penalty, and risk coefficient \(\beta\).  This coefficient converts the
average admitted-task risk constraint into a linear inequality.

\subsection{Risk-Budgeted Assignment}

Let \(\mathcal O_i\) be the retained execution options for task \(i\),
\(\mathcal O=\bigcup_i\mathcal O_i\), and
\(\mathcal D_{\rm edge}=\{1,\ldots,D\}\).
For option \(o=(i,d,\alpha)\), define \(d(o)=d\), \(e_o=e_{id}\),
\(c_o=c_{id}\), and \(\alpha_o=\alpha\).
Let \(C_d>0\) be edge \(d\)'s batch capacity in Gcycles, and
\(\beta\in(0,1)\) the fleet-risk budget.  Binary variables
\(x_o\) and \(z_i\) select an execution option and a drop, respectively.  The offline reference problem is
\begin{align}
 \min_{x,z}\quad&
 \sum_o e_o x_o+M\sum_i z_i, \label{eq:obj}\\
 \text{s.t.}\quad&
 \sum_{o\in\mathcal O_i}x_o+z_i=1,\quad \forall i, \label{eq:assign}\\
 &\sum_{o:d(o)=d}c_o x_o\leq C_d,
 \quad \forall d\in\mathcal D_{\rm edge}, \label{eq:capacity}\\
 &\sum_o\alpha_o x_o\leq
 \beta\sum_i(1-z_i). \label{eq:risk}
\end{align}
Choose \(M>\sum_i\max_{o\in\mathcal O_i}e_o\), with an empty maximum
set to zero.  This exceeds any batch-energy difference, so
the objective first minimizes drops and then minimizes energy.  Constraint
\eqref{eq:assign} chooses one outcome per task.
Constraint~\eqref{eq:capacity} enforces edge compute capacity.
Constraint~\eqref{eq:risk} limits the average nominal risk among admitted
tasks to \(\beta\).  Equivalently,
\(\sum_o\alpha_o x_o+\beta\sum_i z_i\leq\beta N\), giving a drop the
risk coefficient \(\beta\).  This form also covers an all-dropped batch.
The budget controls nominal risk allocation; adaptive reliability is measured
on an independent test stream, as discussed in Section~VII.

\subsection{Repair and Local Improvement}

The online scheduler initializes each task with its least-energy certified
execution option, or a drop if no such option exists.  Let \(a_i\) denote its
selected outcome, including the drop option, with cost
\(\ell(a_i)=e_{a_i}+M\mathbf 1\{a_i\text{ is a drop}\}\).
Define \(\rho(a_i)=\alpha_{a_i}\) for execution and \(\rho(a_i)=\beta\)
for a drop.  For selection \(a=(a_1,\ldots,a_N)\), the total nominal risk and
edge use are
\begin{equation}
 R(a)=\sum_i\rho(a_i),\qquad
 G_d(a)=\sum_{i:d(a_i)=d}c_{a_i}.
 \label{eq:state}
\end{equation}
A drop has destination index \(-1\) and zero edge use.
With \([v]_+=\max(v,0)\), aggregate normalized
violation is
\begin{equation}
 V(a)=\frac{[R(a)-\beta N]_+}{\beta N}
       +\sum_{d=1}^{D}\frac{[G_d(a)-C_d]_+}{C_d}.
 \label{eq:violation}
\end{equation}
For each single-task replacement \(a_i\to b\), let \(a^{i\to b}\) denote the
updated selection and \(\Delta V_{ib}=V(a)-V(a^{i\to b})\).
Among moves with \(\Delta V_{ib}>0\), repair selects
\begin{equation}
 (i^*,b^*)\in\arg\min_{i,b}
 \frac{\ell(b)-\ell(a_i)}{\Delta V_{ib}}.
 \label{eq:repair}
\end{equation}
Ties follow task index and option construction order (local, increasing
remote index, then drop).  Repair stops when \(V(a)=0\), no reducing move
exists, or \(8N\) iterations have run.  Two subsequent coordinate-descent
passes visit tasks in index order and accept the cheapest lower-cost
replacement that keeps \(V=0\).  A pass with no changes ends descent.
The final risk and capacity checks determine whether the schedule is feasible.

Dominance pruning leaves at most \(N(|\mathcal D|+1)\) outcomes, including
drops.  Each repair iteration scans \(O(N|\mathcal D|)\) replacements.
The current implementation recomputes the selection state for each candidate,
giving \(O(N^3|\mathcal D|)\) worst-case work over \(8N\) iterations.
Maintaining risk and capacity totals incrementally would reduce this to
\(O(N^2|\mathcal D|)\) for the same bounded iteration count.

\section{Experimental Setup}

\subsection{Controlled UAV Edge--Cloud Simulator}

We evaluate \method{} in a controlled UAV edge--cloud simulation.
Each primary batch contains 50 UAV tasks, four ground-edge nodes in a
\(1\,\mathrm{km}\times1\,\mathrm{km}\) region, one cloud, and local execution.
Task positions, workload classes, radio conditions, and latent service times
are redrawn for every batch.  All compared methods receive identical tasks and
latent service-time draws within a replicate.  Table~\ref{tab:parameters}
summarizes the primary configuration.

\begin{table}[t]
  \centering
  \caption{Primary simulation parameters.}
  \label{tab:parameters}
  \resizebox{\columnwidth}{!}{
  \begin{tabular}{ll}
    \toprule
    Quantity & Setting \\
    \midrule
    Decision batch & 50 tasks \\
    Region and remote resources & \(1\,\mathrm{km}\times1\,\mathrm{km}\), 4 edges, 1 cloud \\
    Workload input centers & 0.55, 1.9, 4.8 MB \\
    Compute-demand centers & 1.3, 4.8, 10.5 Gcycles \\
    Deadline centers & 0.42, 0.78, 1.38 s \\
    Onboard throughput & 7--11 GHz \\
    Edge throughput & 34--46 GHz \\
    Reporter configuration & \(U=8\), \(m=10{,}000\) per destination \\
    Burst component & 4.5\% of remote scores \\
    Risk grid & \(\{0.01,0.02,0.05,0.10,0.20\}\) \\
    Primary fleet budget & \(\beta=0.05\) \\
    Replication & 30 seeds, 20 batches per condition \\
    \bottomrule
  \end{tabular}}
\end{table}

Tasks follow the three workload profiles in Table~\ref{tab:parameters}.
Edge transmission rate decays with squared distance, and edge compute
throughput varies from 34 to 46 GHz.  Each edge has capacity
\(C_d=38+1.05N/D\) Gcycles.  Cloud energy includes a 0.08-J allocated
gateway and backhaul component; optimized and reported energy use the same
definition.

Remote normalized scores are generated as
\(S=X+JY\), where independent variables satisfy
\(\ln X\sim\mathcal N(-0.35,0.42^2)\),
\(\ln Y\sim\mathcal N(0.75,0.48^2)\), and
\(J\sim\mathrm{Bernoulli}(0.045)\).  The simulator uses \(U=8\) identities and
\(m=10{,}000\) scores per identity and destination.  The primary lure
condition adds 2.8 normalized latency units to honest calibration and test
scores at edge nodes.  Compromised identities report 0.20, and the cloud
remains undegraded.  This construction isolates optimistic calibration from
task generation and preserves an honest clean-score oracle.

All experiments were executed in CPU-only mode on a workstation with one
Intel Xeon E5-2698 v4 processor, 20 physical cores, 40 hardware threads, and
252 GiB of available memory.  The workstation ran 64-bit Linux kernel 5.15.0
and CPython 3.13.12.  Numerical computation used NumPy 2.4.6, SciPy 1.17.1,
and pandas 3.0.3.  Exact assignments used the mixed-integer linear-programming
interface and its HiGHS backend.  Reported scheduler runtimes exclude task
generation, calibration, data serialization, and figure production.

\subsection{Attack Conditions}

Five telemetry conditions separate coverage and availability effects.  The
clean condition leaves all buffers unchanged.  Stealth replaces scores by uniform draws on \([0.28,0.46]\); lure uses
the constant 0.20.  Denial uses uniform draws on \([7,10]\).
For \(k>0\), mixed assigns \(\max(1,\lfloor k/2\rfloor)\) compromised
identities to lure and the remainder to denial.  When \(k>0\), honest edge
calibration and test scores receive shifts of 2.2, 2.8, and 2.5 in stealth,
lure, and mixed conditions, respectively; cloud scores remain unshifted.
The attack conditions therefore pair each report pattern with its stated
service condition.  The experiment
sweeps \(K=0,\ldots,7\) at \(\beta=0.05\).  A separate sensitivity study uses
\(\beta\in\{0.02,0.05,0.10,0.20\}\) at \(K=7\).

Poisoned conditions use \(k=K\); clean conditions use \(k=0\) at every
declared cap.  At \(K=0\), BRC equals pooled conformal.  Clean \(K=7\)
measures the cost of an unused conservative cap.

\subsection{Baselines}

The objective-matched adaptive BRC reference solves
\eqref{eq:obj}--\eqref{eq:risk} over the full risk grid.  The primary
optimality-gap experiment uses zero relative gap tolerance and an 8-s time
limit; the scaling experiment uses a \(10^{-3}\) tolerance and an 8-s limit.
The fixed BRC reference uses only
\(\alpha=\beta\) and solves the same discrete assignment with a \(10^{-3}\)
relative gap tolerance.  A hardened greedy
rule uses fixed BRC bounds but makes deterministic deadline-ordered choices.
These comparisons separate adaptive risk allocation from calibration
robustness and from heuristic optimization.

Ordinary pooled conformal and reporter-median calibration use the same repair
scheduler as \method.  A mean-latency greedy rule represents an uncalibrated
deterministic policy.  A maximum-observation rule replaces calibrated
quantiles by the largest reported score.  The clean-score oracle uses
unpoisoned calibration scores and the same adaptive repair scheduler.  This
oracle measures the price of the certificate without changing the scheduling
logic.

\subsection{Metrics and Statistical Analysis}

The primary evaluation uses \NumSeeds{} independent random seeds and
\NumBatches{} decision batches per seed and condition.  Deadline-miss ratio
(DMR) includes late and dropped tasks.  Energy is normalized by all generated
tasks, including drops.  Runtime measures scheduling only and excludes data
generation and calibration.

Intervals for primary DMR use a clustered bootstrap over seed means with 4,000
resamples.  Energy comparisons are paired by seed because every method receives
common task and service-time draws.  Figure bands and error bars use 95\%
normal intervals over independent seed means.  Scaling uses five seeds and
three batches per task count, giving 15 runtime and completion observations per task count.

\section{Results}

\subsection{Primary Compromise Sweep}

\begin{figure*}[t]
  \centering
  \includegraphics[width=\textwidth]{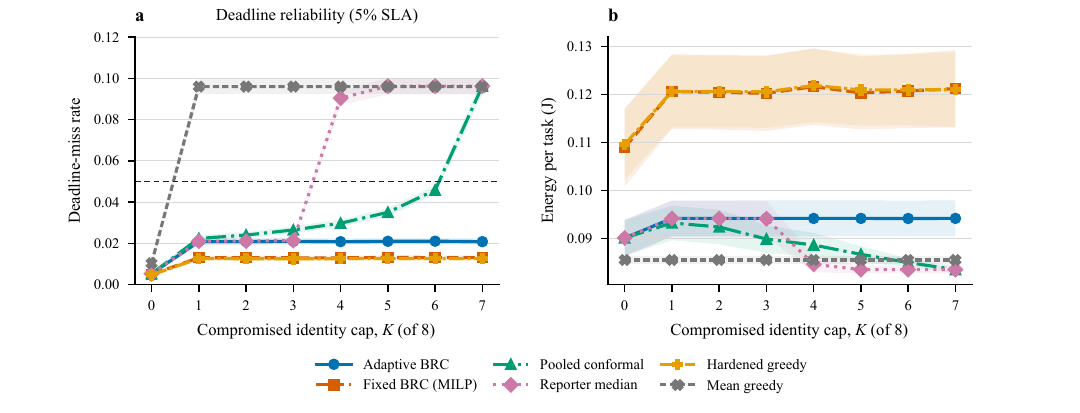}
  \caption{Primary lure-attack sweep over the declared compromised-identity
  cap.  Curves show seed means and shading shows 95\% normal intervals
  (\(n=30\) independent seeds).  (a) DMR with the 5\% service-level threshold.
  (b) End-to-end energy per task.  Energy is normalized by generated tasks, including drops; the fleet-risk
  budget is \(\beta=0.05\).}
  \label{fig:primary}
\end{figure*}

Fig.~\ref{fig:primary} reports the complete lure-attack sweep.  At
\(K\leq3\), reporter-median calibration remains usable because honest
identities retain a majority.  Its DMR rises abruptly at \(K=4\).  Ordinary
pooling degrades more gradually as small scores occupy an increasing share of
the pooled lower ranks.  At \(K=7\), pooled, median, and mean rules reach
\PooledMissPct\%, \MedianMissPct\%, and \MeanMissPct\% DMR.

\method{} remains below the 5\% threshold throughout the sweep.
Its \(K=7\) DMR is \OursMissPct\%, with a 95\% clustered interval of
\OursMissLoPct--\OursMissHiPct\%.  Seven low poisoned buffers consume the
padded ranks, leaving the remaining honest buffer's tail information.

Energy exposes the cost of fixed protection.  At \(K=7\), \method{} consumes
\OursEnergy{} J/task and saves \FixedSavingPct\% relative to fixed-risk BRC
(MILP).
It saves \HardSavingPct\% relative to hardened greedy.  Both comparison rules
remain below the DMR threshold, confirming the energy gain under the same reliability target.  Table~\ref{tab:primary}
reports the corresponding method-level energy, DMR, drop rate, and runtime.

\begin{table}[t]
  \centering
  \caption{Method comparison under lure poisoning at \(k=K=7\) and
  \(\beta=0.05\).  Entries are means over 30 seeds and 20 batches per seed.
  Energy is per generated task; time measures scheduling only.}
  \label{tab:primary}
  \resizebox{\columnwidth}{!}{\input{table_primary.tex}}
\end{table}

\subsection{Risk Sensitivity and Computational Scaling}

\begin{figure*}[t]
  \centering
  \includegraphics[width=\textwidth]{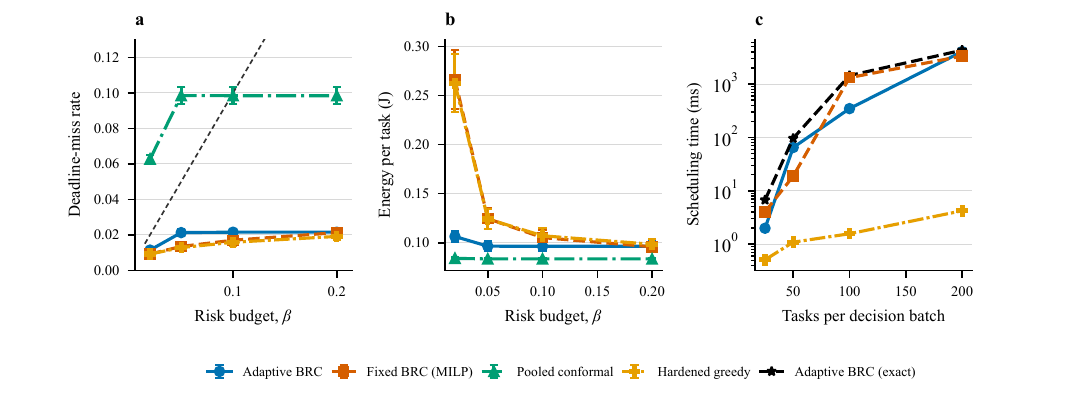}
  \caption{Risk sensitivity and computational scaling under lure poisoning
  at \(k=K=7\).  (a) DMR
  against the nominal fleet budget.  The diagonal marks equality.
  (b) Energy--risk trade-off.  Error bars in (a,b) are 95\% normal intervals
  over seed means (\(n=20\)).  (c) Scheduling time across batch sizes
  (\(n=5\) seeds and three batches per size, logarithmic scale).  Panel (c) shows mean scheduling time over 15 batches per task count.}
  \label{fig:sensitivity}
\end{figure*}

Fig.~\ref{fig:sensitivity}(a,b) varies the fleet budget.  Tightening
\(\beta\) lowers DMR and forces more conservative destinations.  Relaxing
\(\beta\) reduces energy until destination feasibility, rather than aggregate
risk, becomes active.  \method{} stays below the corresponding nominal budget
over the evaluated grid.  Pooled conformal remains unsafe at every displayed
budget under the \(K=7\) lure condition.

Across 25--100 tasks, repair scheduling avoids the adaptive exact assignment's
runtime tail while preserving feasibility.  At 50 tasks, the primary 95th
percentile runtime is \RuntimeP{} ms.  At 200 tasks, repair requires several
seconds and the exact assignment reaches its time limit in some replicates.
The timing results favor repair for the evaluated small and medium batches.

\subsection{Attack Direction and Availability}

\begin{figure*}[t]
  \centering
  \includegraphics[width=\textwidth]{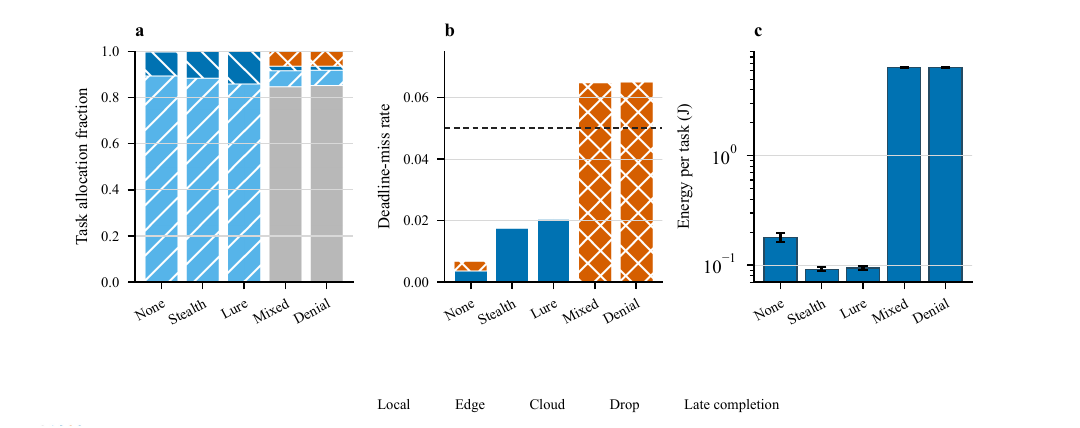}
  \caption{FleetGuard's response to telemetry direction at \(K=7\) and
  \(\beta=0.05\).  Values are means over 30 independent seeds.
  (a) Fractions assigned to local, edge, cloud, and drop outcomes.
  (b) DMR partitioned into late completions and drops; the dashed line marks
  5\%.  (c) Energy per generated task with 95\% normal intervals on a
  logarithmic scale.}
  \label{fig:attacks}
\end{figure*}

Fig.~\ref{fig:attacks} separates execution behavior, failure type, and
energy.  Under clean, stealth, and lure conditions, almost every task executes
at an edge or in the cloud.  Misses are predominantly late completions rather
than admission failures.  The lure condition reaches \OursMissPct\% DMR while
using \OursEnergy{} J/task.  Table~\ref{tab:attacks} summarizes all five
conditions for \method.

\begin{table}[t]
  \centering
  \caption{FleetGuard under five telemetry conditions at \(K=7\) and
  \(\beta=0.05\).  Entries are means over 30 seeds and 20 batches per seed.
  Energy is per generated task.  Clean uses \(k=0\); poisoned conditions use
  \(k=7\).}
  \label{tab:attacks}
  \resizebox{\columnwidth}{!}{\input{table_attacks.tex}}
\end{table}

Mixed and denial reports suppress remote options, shift admitted tasks to
local execution, and cause approximately 6.5\% drops.  Pure denial yields
\DenialMissPct\% DMR, all from drops (\DenialDropPct\%).

The clean condition in Fig.~\ref{fig:attacks} retains the declared \(K=7\)
cap.  BRC selects an extreme pooled quantile and consumes
\UnusedCapEnergy{} J/task.  At \(K=0\), its clean energy overhead relative
to pooled conformal is \CleanOverheadPct\%.

\subsection{Optimality Gap and Runtime}

\begin{figure*}[t]
  \centering
  \includegraphics[width=\textwidth]{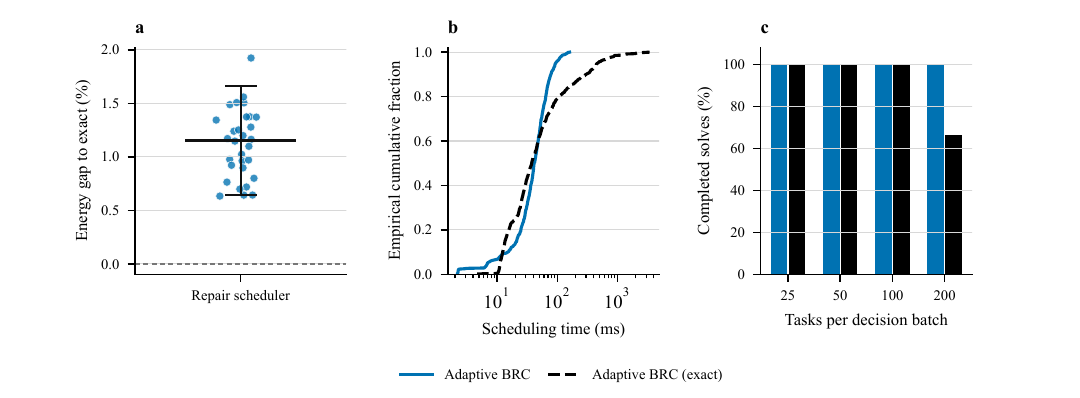}
  \caption{Objective and runtime diagnostics for adaptive BRC under lure
  poisoning at \(k=K=7\) and \(\beta=0.05\).
  (a) Paired seed-level repair energy gap to the exact assignment at 50 tasks
  (\(n=30\)); the bar marks the median and the whisker marks the 2.5th--97.5th
  percentiles.  (b) Empirical runtime distributions over 600 paired batches.
  (c) Feasible repair returns and exact solves reaching the requested
  optimality tolerance within the time limit, over 15 batches per task count.}
  \label{fig:exact}
\end{figure*}

Fig.~\ref{fig:exact}(a) reports the paired seed-level gap to the
zero-tolerance adaptive exact assignment.  Every gap is nonnegative, with a
mean of \ExactGapPct\%, showing that repair approaches the matched optimum.

Exact assignment has a longer runtime tail and a mean runtime
\ExactRuntimeRatio\(\times\) that of repair.  Both complete all 600 primary
batches.  At 200 tasks, all repair schedules are feasible, while two thirds of
MILP runs reach the requested tolerance within 8 s.

\subsection{Robustness Landscape and Oracle Comparison}

\begin{figure*}[t]
  \centering
  \includegraphics[width=\textwidth]{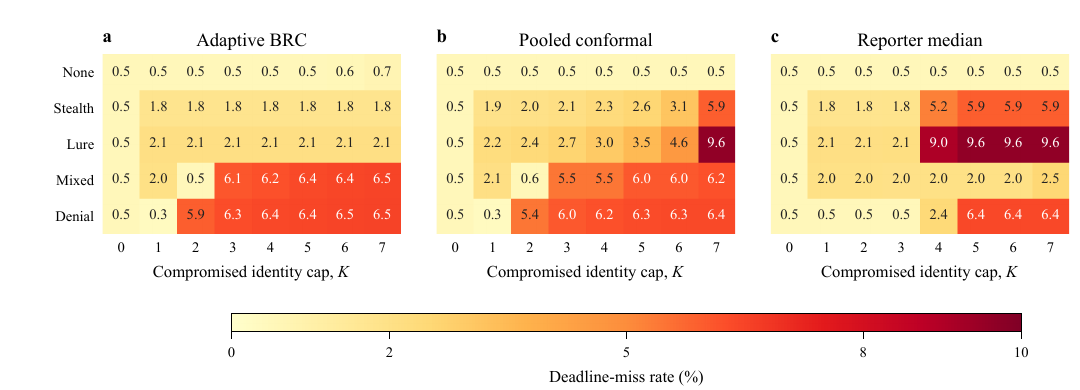}
  \caption{DMR landscape over five telemetry conditions and
  \(K=0,\ldots,7\) at \(\beta=0.05\).  Cells show percentage means over 30
  independent seeds.  (a) Adaptive BRC.  (b) Ordinary pooled conformal.
  (c) Reporter-median calibration.  A common 0--10\% color scale permits
  direct comparison.}
  \label{fig:landscape}
\end{figure*}

Fig.~\ref{fig:landscape} extends the primary lure sweep to every attack
direction.  Adaptive BRC remains below 5\% for clean, stealth, and lure
conditions across the complete \(K\) range.  Its DMR rises under mixed and
denial conditions only after high reports make remote execution unavailable.
At \(K=7\), all mixed and denial misses are drops
(Table~\ref{tab:attacks}).

Pooling degrades under stealth and lure as \(K\) increases.  Reporter-median
calibration exhibits a majority threshold at \(K=4\); its stability under
mixed reports depends on cancellation and supplies no certificate.

The clean-score oracle uses the same repair scheduler and all \(Um\)
unpoisoned scores.  At \(K=7\), its DMR is \OracleMissPct\%, with a
BRC--oracle energy difference of \OracleEnergyGapPct\%.
When \(k=K\) and poisoned scores lie below honest scores, \eqref{eq:rank}
matches the conformal rank of the remaining \((U-K)m\) honest subset.
The experimental oracle has a larger calibration pool; near-matching energy
shows a small empirical protection cost rather than identical quantiles.

The maximum-observation rule consumes \WorstcaseEnergy{} J/task and reaches
\WorstcaseMissPct\% DMR because its excessive bound removes destinations and
forces drops (Table~\ref{tab:primary}).

\section{Discussion}

FleetGuard's energy advantage comes from adaptive risk allocation: the repair
scheduler captures most of the benefit of the matched optimal assignment.
Identity caps protect calibration by reserving ranks for low poisoned scores,
retaining honest tail information without a reporter majority.

High-score poisoning preserves bound conservatism but suppresses remote
options, causing \DenialDropPct\% drops in the denial stress case.
Trusted probes, reserved local capacity, or independently attested edge
measurements could provide a separate availability channel.  An unused
\(K=7\) cap also increases energy; operational \(K\) should follow credential
management and compromise assessment.

The certificate requires authenticated identities, equal raw-score buffers,
and exchangeable honest residuals.  Sybil identities, compromise counts above
\(K\), and persistent drift violate these assumptions.  Short epochs limit
drift exposure at the cost of fewer calibration scores; weighted or online
calibration can address nonstationarity \cite{barber2023beyond}.
Theorem~1 covers levels fixed independently of calibration, while the
scheduler selects levels from the observed curve.  The nominal-risk budget
therefore supplies neither post-selection coverage nor a joint batch
guarantee.  Adaptive reliability is measured empirically; mission-critical
extensions may require simultaneous calibration or deterministic fallback.

Synthetic workloads and residuals require trace replay, hardware-in-the-loop
tests, and flight validation to establish platform-specific behavior.
Cross-condition differences reflect both report direction and modeled service
degradation.  Repair reaches seconds at 200 tasks and can terminate without
feasibility outside the evaluated cases.  Incremental state updates and
edge-wise decomposition should be assessed against the same feasibility and
energy criteria for larger batches.

\section{Conclusion}

FleetGuard combines identity-capped calibration and adaptive risk allocation
for UAV offloading.  Under seven-reporter lure poisoning, it achieves
\OursMissPct\% DMR and \FixedSavingPct\% energy savings over fixed BRC,
with a \ExactGapPct\% gap to adaptive exact assignment and
\ExactRuntimeRatio\(\times\) lower mean scheduling time.

\bibliographystyle{IEEEtran}
\bibliography{refs}

\end{document}

%% file: results_macros.tex
% Generated by scripts/make_paper_assets.py; do not edit by hand.
\newcommand{\OursMissPct}{2.08}
\newcommand{\OursMissLoPct}{1.94}
\newcommand{\OursMissHiPct}{2.24}
\newcommand{\OursEnergy}{0.094}

\newcommand{\FixedSavingPct}{22.3}
\newcommand{\HardSavingPct}{22.2}

\newcommand{\PooledMissPct}{9.62}
\newcommand{\MedianMissPct}{9.63}
\newcommand{\MeanMissPct}{9.60}
\newcommand{\RuntimeP}{94.0}
\newcommand{\ExactGapPct}{1.1}
\newcommand{\ExactRuntimeRatio}{2.6}
\newcommand{\CleanOverheadPct}{0.0}
\newcommand{\UnusedCapEnergy}{0.180}
\newcommand{\OracleEnergyGapPct}{0.02}
\newcommand{\OracleMissPct}{2.09}
\newcommand{\WorstcaseEnergy}{6.472}
\newcommand{\WorstcaseMissPct}{6.38}
\newcommand{\DenialMissPct}{6.49}
\newcommand{\DenialDropPct}{6.49}
\newcommand{\NumSeeds}{30}
\newcommand{\NumBatches}{20}

%% file: table_primary.tex
% Generated by scripts/make_paper_assets.py.
\begin{tabular}{lcccc}
\toprule
Method & Energy (J/task) & DMR (\%) & Drop (\%) & Time (ms) \\
\midrule
FleetGuard & 0.094 & 2.08 & 0.03 & 45.7 \\
FleetGuard-exact & 0.093 & 2.08 & 0.03 & 118.7 \\
Fixed BRC (MILP) & 0.121 & 1.31 & 0.09 & 51.7 \\
Hardened greedy & 0.121 & 1.26 & 0.09 & 1.7 \\
Pooled conformal & 0.084 & 9.62 & 0.00 & 170.9 \\
Reporter median & 0.084 & 9.63 & 0.00 & 170.4 \\
Mean greedy & 0.086 & 9.60 & 0.00 & 1.6 \\
Clean-score oracle & 0.094 & 2.09 & 0.03 & 77.2 \\
Maximum observation & 6.472 & 6.38 & 6.38 & 3.9 \\
\bottomrule
\end{tabular}

%% file: table_attacks.tex
% Generated by scripts/make_paper_assets.py.
\begin{tabular}{lccc}
\toprule
Telemetry condition & Energy (J/task) & DMR (\%) & Drop (\%) \\
\midrule
No poisoning & 0.180 & 0.71 & 0.32 \\
Lure (low reports) & 0.094 & 2.08 & 0.03 \\
Stealthy low reports & 0.092 & 1.79 & 0.03 \\
Mixed low/high & 6.411 & 6.47 & 6.47 \\
Denial (high reports) & 6.445 & 6.49 & 6.49 \\
\bottomrule
\end{tabular}

%% file: refs.bib
@article{angelopoulos2023gentle,
  author  = {Anastasios N. Angelopoulos and Stephen Bates},
  title   = {Conformal Prediction: A Gentle Introduction},
  journal = {Foundations and Trends in Machine Learning},
  volume  = {16},
  number  = {4},
  pages   = {494--591},
  year    = {2023},
  doi     = {10.1561/2200000101}
}

@article{barber2023beyond,
  author  = {Rina Foygel Barber and Emmanuel J. Cand{\`e}s and Aaditya Ramdas and Ryan J. Tibshirani},
  title   = {Conformal Prediction Beyond Exchangeability},
  journal = {Annals of Statistics},
  volume  = {51},
  number  = {2},
  pages   = {816--845},
  year    = {2023},
  doi     = {10.1214/23-AOS2276}
}

@inproceedings{kang2024robfcp,
  author    = {Mintong Kang and Zhen Lin and Jimeng Sun and Cao Xiao and Bo Li},
  title     = {Certifiably {B}yzantine-Robust Federated Conformal Prediction},
  booktitle = {Proc. 41st Int. Conf. Machine Learning (ICML)},
  series    = {PMLR},
  volume    = {235},
  pages     = {23022--23057},
  year      = {2024}
}

@article{lari2026partial,
  author  = {Ehsan Lari and Reza Arablouei and Stefan Werner},
  title   = {Partial Model Sharing Improves {B}yzantine Resilience in Federated Conformal Prediction},
  journal = {arXiv preprint arXiv:2605.11684},
  year    = {2026}
}

@inproceedings{li2024poison,
  author    = {Yangyi Li and Aobo Chen and Wei Qian and Chenxu Zhao and Divya Lidder and Mengdi Huai},
  title     = {Data Poisoning Attacks against Conformal Prediction},
  booktitle = {Proc. 41st Int. Conf. Machine Learning (ICML)},
  series    = {PMLR},
  volume    = {235},
  pages     = {27563--27574},
  year      = {2024}
}

@inproceedings{scholten2025rps,
  author    = {Yan Scholten and Stephan G{\"u}nnemann},
  title     = {Provably Reliable Conformal Prediction Sets in the Presence of Data Poisoning},
  booktitle = {Proc. Int. Conf. Learning Representations (ICLR)},
  year      = {2025}
}

@inproceedings{park2023acon2,
  author    = {Sangdon Park and Osbert Bastani and Taesoo Kim},
  title     = {{ACon}$^2$: Adaptive Conformal Consensus for Provable Blockchain Oracles},
  booktitle = {Proc. 32nd USENIX Security Symp.},
  pages     = {3313--3330},
  year      = {2023}
}

@article{einbinder2024noise,
  author  = {Bat-Sheva Einbinder and Shai Feldman and Stephen Bates and Anastasios N. Angelopoulos and Asaf Gendler and Yaniv Romano},
  title   = {Label Noise Robustness of Conformal Prediction},
  journal = {Journal of Machine Learning Research},
  volume  = {25},
  number  = {328},
  pages   = {1--66},
  year    = {2024}
}

@article{guo2024multiuav,
  author  = {Hongzhi Guo and Yutao Wang and Jiajia Liu and Chang Liu},
  title   = {Multi-{UAV} Cooperative Task Offloading and Resource Allocation in {5G} Advanced and Beyond},
  journal = {IEEE Transactions on Wireless Communications},
  volume  = {23},
  number  = {1},
  pages   = {347--359},
  year    = {2024},
  doi     = {10.1109/TWC.2023.3277801}
}

@article{yang2020loadbalance,
  author  = {Lei Yang and Haipeng Yao and Jingjing Wang and Chunxiao Jiang and Abderrahim Benslimane and Yunjie Liu},
  title   = {Multi-{UAV}-Enabled Load-Balance Mobile-Edge Computing for {IoT} Networks},
  journal = {IEEE Internet of Things Journal},
  volume  = {7},
  number  = {8},
  pages   = {6898--6908},
  year    = {2020}
}

@article{hui2024altitude,
  author  = {Min Hui and Jian Chen and Long Yang and Lu Lv and Hai Jiang and Naofal Al-Dhahir},
  title   = {{UAV}-Assisted Mobile Edge Computing: Optimal Design of {UAV} Altitude and Task Offloading},
  journal = {IEEE Transactions on Wireless Communications},
  volume  = {23},
  number  = {10},
  pages   = {13633--13647},
  year    = {2024},
  doi     = {10.1109/TWC.2024.3403536}
}

@article{xu2024trajectory,
  author  = {Wenlong Xu and Tiankui Zhang and Xidong Mu and Yuanwei Liu and Yapeng Wang},
  title   = {Trajectory Planning and Resource Allocation for Multi-{UAV} Cooperative Computation},
  journal = {IEEE Transactions on Communications},
  volume  = {72},
  number  = {7},
  pages   = {4305--4318},
  year    = {2024},
  doi     = {10.1109/TCOMM.2024.3361536}
}

@inproceedings{chen2023emergency,
  author    = {Chaobin Chen and Tiankui Zhang and Wenjun Xu and Xu Yang and Yapeng Wang},
  title     = {Multi-{UAV} Cooperation Based Edge Computing Offloading in Emergency Communication Networks},
  booktitle = {Proc. IEEE Wireless Communications and Networking Conf. (WCNC)},
  pages     = {1--6},
  year      = {2023},
  doi       = {10.1109/WCNC55385.2023.10118864}
}

@article{liu2023federated,
  author  = {Tianze Liu and Tian Kui Zhang and Jonathan Loo and Ya Peng Wang},
  title   = {Deep Reinforcement Learning-Based Resource Allocation for {UAV}-Enabled Federated Edge Learning},
  journal = {Journal of Communications and Information Networks},
  volume  = {8},
  number  = {1},
  pages   = {1--12},
  year    = {2023},
  doi     = {10.23919/JCIN.2023.10087243}
}

@inproceedings{wei2024path,
  author    = {Junhao Wei and Yanzhao Gu and K. L. Eddie Law and Ngai Cheong},
  title     = {Adaptive Position Updating Particle Swarm Optimization for {UAV} Path Planning},
  booktitle = {Proc. 22nd Int. Symp. Modeling and Optimization in Mobile, Ad Hoc, and Wireless Networks (WiOpt)},
  pages     = {124--131},
  year      = {2024}
}

@article{wei2026aerotsboost,
  author  = {Junhao Wei and Haochen Li and Yanxiao Li and Yifu Zhao and Dexing Yao and Baili Lu and Xudong Ye and Sio-Kei Im and Yapeng Wang and Xu Yang},
  title   = {{AeroTSBoost}: Temporal-Statistical Boosting for Real-World {UAV} Telemetry Anomaly Mining},
  journal = {arXiv preprint arXiv:2605.25639},
  year    = {2026},
  doi     = {10.48550/arXiv.2605.25639}
}

@article{liu2024jamming,
  author  = {Shuai Liu and Helin Yang and Liang Xiao and Mengting Zheng and Huabing Lu and Zehui Xiong},
  title   = {Learning-Based Resource Management Optimization for {UAV}-Assisted {MEC} Against Jamming},
  journal = {IEEE Transactions on Communications},
  volume  = {72},
  number  = {8},
  pages   = {4873--4886},
  year    = {2024},
  doi     = {10.1109/TCOMM.2024.3374356}
}

@article{cui2025robust,
  author  = {Can Cui and Ziye Jia and Chao Dong and Qihui Wu},
  title   = {Robust Task Offloading for {UAV}-Enabled Secure {MEC} Against Aerial Eavesdropper},
  journal = {arXiv preprint arXiv:2507.00710},
  year    = {2025}
}

@inproceedings{zhang2019dedas,
  author    = {Jiaying Meng and Haisheng Tan and Chao Xu and Wanli Cao and Liuyan Liu and Bojie Li},
  title     = {Dedas: Online Task Dispatching and Scheduling with Bandwidth Constraint in Edge Computing},
  booktitle = {Proc. IEEE INFOCOM},
  pages     = {2287--2295},
  year      = {2019},
  doi       = {10.1109/INFOCOM.2019.8737577}
}

@article{wei2026geossa,
      title={{GeoSSA}: Geometric Sparrow Search Algorithm for {UAV} Path Planning and Engineering Design Optimization}, 
      author={Junhao Wei and Wenxuan Zhu and Qingyang Xu and Yanxiao Li and Yifu Zhao and Zikun Li and Ran Zhang and Yanzhao Gu and Jinhong Song and Yapeng Wang and Zhiwen Wang and Ngai Cheong and Sio-Kei Im and Xu Yang},
      year={2026},
      journal={arXiv preprint arXiv:2601.19346},
      doi={10.48550/arXiv.2601.19346},
      url={https://arxiv.org/abs/2601.19346}, 
}

@article{wei2026saga,
      title={{SAGA}: A Robust Self-Attention and Goal-Aware Anchor-based Planner for Safe {UAV} Autonomous Navigation}, 
      author={Junhao Wei and Yanxiao Li and Dexing Yao and Yifu Zhao and Haochen Li and Qibin He and Baili Lu and Xiaofan Zou and Dingcheng Yang and Sio-Kei Im and Yapeng Wang and Xu Yang},
      year={2026},
      journal={arXiv preprint arXiv:2605.02301},
      doi={10.48550/arXiv.2605.02301},
      url={https://arxiv.org/abs/2605.02301}, 
}

@article{wei2026labhh,
      title={Landscape-Aware Bandit Hyper-Heuristics for Online Operator Selection in {UAV} Inspection Routing}, 
      author={Junhao Wei and Yanxiao Li and Yifu Zhao and Qibin He and Haochen Li and Dexing Yao and Baili Lu and Zhenhong Peng and Yapeng Wang and Sio-Kei Im and Xu Yang},
      year={2026},
      journal={arXiv preprint arXiv:2605.14620},
      doi={10.48550/arXiv.2605.14620},
      url={https://arxiv.org/abs/2605.14620}, 
}

@article{wei2026aries2,
      title={{ARIES-Mission2}: A Zero-Shot Vision-Language-Action Framework for Fast Large-Scale Aerial Mission Generation}, 
      author={Junhao Wei and Yanxiao Li and Haochen Li and Yifu Zhao and Dexing Yao and Baili Lu and Zikun Li and Yapeng Wang and Sio-Kei Im and Dingcheng Yang and Xu Yang},
      year={2026},
      journal={arXiv preprint arXiv:2608.12763},
      doi={10.48550/arXiv.2608.12763},
      url={https://arxiv.org/abs/2608.12763}, 
}

@article{Li2026RAWARHP,
  title   = {{RAWA-RHP}: Residual-State Constrained Insertion for Wind-Affected {UAV} Infrastructure Inspection},
  author  = {Li, Haochen and Wei, Junhao and Yao, Dexing and Zhao, Yifu and Li, Yanxiao and Mai, Zhaoyan and Yu, Jiamu and Lu, Baili and Im, Sio-Kei and Yang, Xu and Wang, Yapeng},
  journal = {Results in Engineering},
  year    = {2026},
  pages   = {113481},
  doi     = {10.1016/j.rineng.2026.113481},
  url     = {https://doi.org/10.1016/j.rineng.2026.113481}
}

@misc{Wei2026ARIESMission,
  title     = {{ARIES-Mission}: A Vision-Language System for Efficient {UAV} Mission Generation},
  author    = {Wei, Junhao and Li, Haochen and Zhao, Yifu and Yao, Dexing and Li, Yanxiao and Lu, Baili and Im, Sio-Kei and Yang, Xu and Wang, Yapeng},
  year      = {2026},
  publisher = {Research Square},
  doi       = {10.21203/rs.3.rs-10372117/v1},
  url       = {https://doi.org/10.21203/rs.3.rs-10372117/v1},
  note      = {Preprint}
}
